\documentclass[conference]{IEEEtran}
\usepackage{cite}
\usepackage{amsmath,amssymb,amsfonts}
\usepackage{algorithmic}
\usepackage{graphicx, ifthen}
\usepackage{subfig}
\usepackage{tikz, pgfplots, pgfmath}
\usetikzlibrary{positioning, fit, calc, decorations.markings}
\usepackage{textcomp}
\usepackage{enumerate}
\usepackage{xcolor}
\usepackage{braket}
\usepackage{physics}
\usepackage{comment}
\usepackage{mathtools}
\usepackage{hyperref}
\usetikzlibrary{3d}
\usepackage{tikz-3dplot}

\usepackage[stretch=15,shrink=20]{microtype}

\def\BibTeX{{\rm B\kern-.05em{\sc i\kern-.025em b}\kern-.08em
    T\kern-.1667em\lower.7ex\hbox{E}\kern-.125emX}}

\usepackage[utf8]{inputenc}
\usepackage[english]{babel}
\usepackage[T1]{fontenc}

\usepackage{lipsum}

\usepackage{makecell}
\setcellgapes{4pt}

\usepackage{algorithm}

\IEEEoverridecommandlockouts

\begin{document}

\title{Temporal Framework for Causality-Preserving Scheduling of Measurements in Quantum Networks} 

\author{\IEEEauthorblockN{Jakob Kaltoft S{\o}ndergaard}
\IEEEauthorblockA{\textit{Department of Electronic Systems} \\
\textit{Aalborg University}\\
Aalborg, Denmark \\
jakobks@es.aau.dk}
\and
\IEEEauthorblockN{Ren{\'e} B{\o}dker Christensen}
\IEEEauthorblockA{\textit{Department of Mathematical Sciences} \\
\textit{Aalborg University}\\
Aalborg, Denmark \\
rene@math.aau.dk}
\and
\IEEEauthorblockN{Petar Popovski}
\IEEEauthorblockA{\textit{Department of Electronic Systems} \\
\textit{Aalborg University}\\
Aalborg, Denmark \\
petarp@es.aau.dk}
\thanks{This work was supported, in part, by the Danish National Research Foundation (DNRF), through the Center CLASSIQUE, grant nr. 187.}
}

\maketitle

\begin{abstract}
Distributed quantum protocols rely on classical feedforward information to process measurement outcomes, but heterogeneous hardware and uncertain local timing can make the causal order of measurements ambiguous when inferred solely from arrival times. Even in simple line networks with only Pauli measurements, end nodes cannot distinguish whether a missing outcome is caused by slow measurement or by delayed classical propagation. To resolve this ambiguity, we propose a time-division architecture for quantum networks in which nodes perform measurements in pre-assigned slots, ensuring a unique causal interpretation of outcomes. We formalize this temporal framework and derive the feedforward and adjacency constraints required to preserve measurement causality. For simple network topologies, we present an algorithm that yields optimal measurement schedules. Overall, the proposed time-division model provides a practical coordination layer that bridges the classical network timing with quantum measurement processing, enabling reliable and scalable measurement-based quantum networking.
\end{abstract}


\section{Introduction}
Quantum networks \cite{wehner2018quantum} are envisioned as the enabling infrastructure for future quantum applications including communication, distributed computing, and sensing. In these networks, entanglement shared between spatially separated nodes is the crucial resource that enables quantum information to be shared and processed across distance. Therefore, entanglement distribution and processing form the backbone of quantum networks. As research progresses from point-to-point links to multi-node architectures, classical \emph{coordination} of entanglement processing between separated nodes becomes increasingly important. Crucial factors for coordination include timing, ordering, and interpretation of the classical messages that accompany quantum operations. The increasing significance of coordination resembles the evolution of classical networking, where scalability and reliability of communication ultimately depends as much on synchronization and scheduling as on the physical channel itself. For example, asynchronous multiple-access channels suffer from uncertainty of event ordering and destructive interference (\emph{collisions}) when transmissions overlap \cite{falconer2002time}. The Time-Division Multiple Access (TDMA) protocol allocates discrete time slots for each node in which they can transmit over the same frequency channel. This prevents collisions by mapping physical network constraints into a predictive temporal structure. In quantum networks, coordination is generally overlooked as a performance measure in favor of, for example, end-to-end fidelity, entanglement rate, or latency.

Measurement-Based Quantum Networking (MBQN) provides a promising architecture \cite{pirker2019quantum} for fast entanglement distribution. In this framework, multipartite entanglement is shared among the nodes prior to an entanglement request. As the pre-shared entanglement serves as the quantum resource for the network, it is commonly referred to as the \emph{resource state}.
When some form of entanglement is requested from the network, henceforth referred to as a \emph{task}, the resource state is altered through a sequence of local measurements. The measurements are performed such that the resulting state is equivalent to the task up to local Clifford operations that depends on the measurement outcomes. Thus, once the resource state has been distributed, MBQN can satisfy a task solely with local operations and classical communication (LOCC). As a result, MBQN supports fast entanglement distribution over long distances by not relying on repeated probabilistic entanglement generation and swapping. Because the measurement sequence is determined after task arrival, MBQN offers high flexibility in the tasks that can be satisfied from a given resource state. Note that the resource state can be designed such that it can satisfy any permitted task.
Prior work has primarily focused on the ``quantum layer''; designing the resource state \cite{miguel2023optimized, sondergaard2025satellite} or deciding the optimal measurement sequence \cite{mor2025imperfect, freund2024flexible}. What remains unexplored is the \emph{coordination layer}, dealing with temporal schedule of the network nodes and interpretation of their measurements.

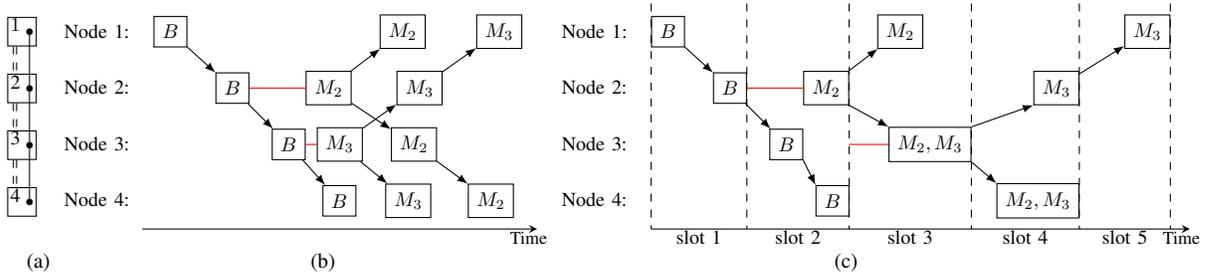
\begin{figure*}[t]
    \centering
    \subfloat[\label{fig: Introduction_Example_Topology}]{%
    \begin{minipage}{.05\textwidth}
    \centering
    \tikzset{qubit/.style={circle, black, draw, fill, very thin, inner sep=1pt},
user/.style={rectangle, draw, black, inner sep=.15cm}
}

\begin{tikzpicture}[scale=.75, transform shape, baseline=(bb.north)]
    \path[use as bounding box] (0,0) coordinate (bb) (-.25,-3.7) rectangle (.25,0.2);

    \node[qubit] at (0,0) (1) {};
    \node[above left=0pt of 1, yshift=-4.5pt] {$1$};
    \node[user, fit=(1), xshift=-.10cm] (U1) {};

    \foreach \n in {2,...,4}{
         \pgfmathtruncatemacro{\nprior}{\n-1}
         \node[qubit, below=.9cm of \nprior] (\n) {};
         \node[above left=0pt of \n, yshift=-4.5pt] {\n};
         \node[user, fit=(\n), xshift=-.10cm] (U\n) {};
         \draw (\nprior) -- (\n);
         \draw[dashed] (U\nprior.240) -- (U\n.120);
         \draw[dashed] (U\nprior.250) -- (U\n.110);
    }

\end{tikzpicture}
    \end{minipage}
    }
    \hspace{.02\textwidth}
    \subfloat[\label{fig: Introduction_Example_Measurement_Interpretation_Asynchronuos}]{%
    \begin{minipage}{.3\textwidth}
    \centering
    \tikzset{cc/.style={rectangle, draw, inner sep=.15cm}}

\begin{tikzpicture}[scale=.75, transform shape, baseline=(bb.south)]
\path[use as bounding box] (0,0) coordinate (bb) (-1,-3.7) rectangle (7,0.2);

\node[cc] at (0,0) (1a) {$B$};
\node[cc, right=.5cm of 1a, yshift=-1cm] (2a) {$B$};
\node[cc, right=.4cm of 2a, yshift=-1cm] (3a) {$B$};
\node[cc, right=.3cm of 3a, yshift=-1cm] (40) {$B$};

\node[cc, right=1cm of 2a] (2b) {$M_2$};
\node[cc, right=.5cm of 2b, yshift=1cm] (1b) {$M_2$};
\node[cc, right=.7cm of 2b, yshift=-1cm] (3c) {$M_2$};
\node[cc, right=.55cm of 3c, yshift=-1cm] (4b) {$M_2$};

\node[cc, right=.2cm of 3a] (3b) {$M_3$};
\node[cc, right=.6cm of 3b, yshift=1cm] (2c) {$M_3$};
\node[cc, right=.6cm of 2c, yshift=1cm] (1c) {$M_3$};
\node[cc, right=.4cm of 3b, yshift=-1cm] (4a) {$M_3$};

\draw[-latex] (1a) -- (2a);
\draw[-latex] (2a) -- (3a);
\draw[-latex] (2b) -- (1b);
\draw[-latex] (2b) -- (3c);
\draw[-latex] (2c) -- (1c);
\draw[-latex] (3a) -- (40);
\draw[-latex] (3b) -- (2c);
\draw[-latex] (3b) -- (4a);
\draw[-latex] (3c) -- (4b);

\draw[red] (2a) -- (2b);
\draw[red] (3a) -- (3b);

\node[left=.3cm of 1a] (01) {Node 1:};
\node[below=.5cm of 01] (02) {Node 2:};
\node[below=.5cm of 02] (03) {Node 3:};
\node[below=.5cm of 03] (04) {Node 4:};

\draw[-stealth] (-0.5,-3.5) -- (6.5,-3.5);
\node[] at (6.35,-3.65) {\small Time};

\end{tikzpicture}
    \end{minipage}
    }
    \hspace{.02\textwidth}
    \subfloat[\label{fig: Introduction_Example_Measurement_Interpretation_Slotted}]{%
    \begin{minipage}{.4\textwidth}
    \centering
    \tikzset{cc/.style={rectangle, draw, inner sep=.15cm}}

\begin{tikzpicture}[scale=.75, transform shape, baseline=(bb.south)]
\path[use as bounding box] (0,0) coordinate (bb) (-1,-3.7) rectangle (7,0.2);

\node[cc] at (0,0) (1a) {$B$};
\node[cc, right=.5cm of 1a, yshift=-1cm] (2a) {$B$};
\node[cc, right=.4cm of 2a, yshift=-1cm] (3a) {$B$};
\node[cc, right=.22cm of 3a, yshift=-1cm] (40) {$B$};

\node[cc, right=1cm of 2a] (2b) {$M_2$};
\node[cc, right=.5cm of 2b, yshift=1cm] (1b) {$M_2$};
\node[cc, right=.7cm of 2b, yshift=-1cm] (3b) {$M_2,M_3$};
\node[cc, right=1.1cm of 3c, yshift=-1cm] (4a) {$M_2,M_3$};

\node[cc, right=1.1cm of 3b, yshift=1cm] (2c) {$M_3$};
\node[cc, right=.8cm of 2c, yshift=1cm] (1c) {$M_3$};

\draw[-latex] (1a) -- (2a);
\draw[-latex] (2a) -- (3a);
\draw[-latex] (2b) -- (1b);
\draw[-latex] (2b.south east) -- (3b.north west);
\draw[-latex] (2c) -- (1c);
\draw[-latex] (3a.south east) -- (40.north west);
\draw[-latex] (3b.north east) -- (2c.south west);
\draw[-latex] (3b.south east) -- (4a.north west);

\draw[red] (2a) -- (2b);
\draw[red] ($(2b) + (.4,-1)$) -- (3b);

\node[left=.3cm of 1a] (01) {Node 1:};
\node[below=.5cm of 01] (02) {Node 2:};
\node[below=.5cm of 02] (03) {Node 3:};
\node[below=.5cm of 03] (04) {Node 4:};

\draw[dashed] ($(1a.west)+(0,.5)$) -- ($(1a.west)+(0,-3.5)$);
\draw[dashed] ($(2a.east)+(0,1.5)$) -- ($(2a.east)+(0,-2.5)$);
\draw[dashed] ($(2b.east)+(0,1.5)$) -- ($(2b.east)+(0,-2.5)$);
\draw[dashed] ($(3b.east)+(0,2.5)$) -- ($(3b.east)+(0,-1.5)$);
\draw[dashed] ($(4a.east)+(0,3.5)$) -- ($(4a.east)+(0,-.5)$);
\draw[dashed] ($(1c.east)+(0,.5)$) -- ($(1c.east)+(0,-3.5)$);

\node[yshift=-3.15cm] at ($(1a.west)!0.5!(2a.east)$) {slot 1};
\node[yshift=-2.65cm] at ($(2a.east)!0.5!(2b.east)$) {slot 2};
\node[yshift=-2.15cm] at ($(2b.east)!0.5!(3b.east)$) {slot 3};
\node[yshift=-1.15cm] at ($(3b.east)!0.5!(4a.east)$) {slot 4};
\node[yshift=-3.15cm] at ($(2c.east)!0.5!(1c.east)$) {slot 5};

\draw[-stealth] (-0.3,-3.5) -- (9.25,-3.5);
\node[] at (9.1,-3.65) {\small Time};

\end{tikzpicture}
    \end{minipage}
    }
    \caption{Four-node network illustrating measurement causality ambiguity. (a) Topology with initial resource state and single-hop communication channels. (b) Asynchronous measurements: heterogeneous durations cause end nodes to see different measurement orders. (c) Time-slotted measurements enforce a shared causal structure, enabling consistent interpretation of outcomes.}
    \label{fig: QCNC_Introductoy_Example}
\end{figure*}

Consider the simple topology in Fig.~\ref{fig: Introduction_Example_Topology}, where the objective is for nodes~1 and 4 to share a requested entanglement that can be achieved by a sequence of Pauli measurements on nodes~2 and 3. Node~1, acting as the initiator, first computes the required measurement bases for the nodes and encodes this information into a classical message $B$, which is propagated hop-wise toward node~4. When $B$ reaches the nodes~2 and 3, each performs its prescribed measurement and sends the outcome ($M_i$ for node $i$) back toward the end nodes for final corrections, again hop-wise.
In a heterogeneous network, both the measurement durations and the classical forwarding delays vary unpredictably. As a result, the outcome from node~2 may reach node~1 before the outcome from node~3, while the reverse ordering may be observed at node~4 as illustrated in Fig.~\ref{fig: Introduction_Example_Measurement_Interpretation_Asynchronuos}.
Although the end nodes receive both measurement outcomes, they cannot consistently infer which measurement actually completed first, inducing a causality ambiguity. Yet, the relative order is essential; it determines which final corrections to apply at the end nodes. If the end nodes disagree, they apply non-compatible corrections resulting in entanglement that differs from the requested. Note that the order cannot even be reconstructed by buffering until all outcomes are received as the outcomes contain no information about the causality. One might imagine resolving this ambiguity by having each node timestamp its measurement using a Lamport logical clock \cite{lamport2019time}. While the hop-wise transmission of $B$ establishes a causal order for feedforward arrival, it does not enforce a corresponding order of measurement completions; heterogeneous hardware may cause node~3 to complete its measurement before node~2, even though it received $B$ later. Since the measurements themselves are local and concurrent events with no causal messages between them, logical clocks cannot impose a consistent measurement order. Although causality could, in principle, be enforced through explicit post-measurement signaling, such coordination introduces additional latency, requires global synchronization, and is ill-suited as an architectural primitive for scalable quantum networks.

We therefore resolve the ambiguity by assigning each measurement to a discrete time slot, thereby enforcing a consistent causal structure as illustrated in Fig.~\ref{fig: Introduction_Example_Measurement_Interpretation_Slotted}. This ensures that end nodes interpret outcomes in the same logical order, independent of heterogeneous measurement durations and communication delays.

Whether an MBQN can afford a global time-slotted structure depends on three physical timescales: the propagation time of classical feedforward, the duration of quantum measurements including local processing, and the coherence time for the qubits in the resource state. If coherence times are short, nodes must measure almost immediately upon receiving feedforward. In this regime, if the measurement time of a node exceeds the feedforward propagation delay, no coordination mechanism can guarantee a \emph{causality-preserving} measurement schedule. For example, if node~3 in Fig.~\ref{fig: QCNC_Introductoy_Example} cannot delay its measurement without risking its qubit to decohere, the end nodes cannot reliably infer the measurement order. In the opposite extreme, when coherence times exceeds the propagation delays, nodes can defer measurements until prescribed slots. In this regime, time-slotting becomes not only feasible, but a simple, robust, and structured mechanism for preserving measurement causality and avoiding outcome misinterpretation. Time-slotting can therefore be viewed as an architectural design principle for future MBQNs operating with higher coherence times, where temporal coordination is technologically possible and architectural beneficial. In coherence-limited settings, time-slotting cannot be imposed without risking decoherence, and causal ambiguity becomes an inherent limitation of the network rather than a design choice.

Beyond the illustrative four-node scenario, coordination is intrinsic to MBQNs. In general MBQN, where arbitrary entanglements may be requested, preserving the causal ordering of measurements is essential, as the correctness of the generated entanglement depends explicitly on the measurement order. Even when restricting to Pauli measurements, for which the final quantum state is invariant to measurement order, coordination remains necessary to ensure consistent interpretation of outcomes across the network. Heterogeneous measurement durations and asynchronous communication may otherwise cause different nodes to infer incompatible causal orders despite receiving identical sets of outcomes.

These observations motivate imposing a time-division model as an architectural coordination layer for MBQNs. Drawing analogy to TDMA, nodes perform measurements on the shared resource state only at prescribed time slots, ensuring a shared and consistent causal structure. Without such temporal structure, end nodes cannot distinguish whether missing feedback is due to delayed measurements, communication latency, or packet loss, preventing reliable online processing. By contrast, time-slotting enables \emph{streaming} operations: end nodes can process outcomes incrementally and apply corrections as soon as all prerequisite measurements are known. This reduces latency, lowers classical memory requirements, and provides robustness against heterogeneous hardware and communication delays.

The remainder of the paper develops a formal model for time-division coordination in MBQNs. Sec.~\ref{sec: Model} introduces the necessary preliminaries on MBQN, defines the temporal framework, and formalizes causality-preserving measurement schedules. Sec.~\ref{sec: Extreme_Cases} and Sec.~\ref{sec: Scheduling_Heuristics} illustrate its behavior in limiting and general regimes. Sec.~\ref{sec: Results} utilizes the model in simulation, and Sec.~\ref{sec: Conclusion} summarizes the contributions.

\section{Model}\label{sec: Model}
In this section, we formalize the temporal framework. First, we briefly introduce graph states as they are the building blocks of MBQNs. Second, we present the considered network model and the temporal framework. Third, we define causality-preserving measurement schedules.

\subsection{Graph states}
Graph states \cite{hein2006entanglement} is a class of multipartite entangled states that can conveniently be described by a simple graph. Let $G=(V,E)$ denote a simple graph; each vertex represents a qubit, while each edge represents an entangling operation. The graph state defined by $G$ is given as
\begin{equation}
    \ket{G}=\prod_{(a,b)\in E}\text{CZ}_{a,b}\ket{+}^{\otimes\abs{V}},
\end{equation}
where $\text{CZ}_{a,b}$ is the controlled-phase operator applied on qubits $a\otimes b$, which has matrix representation $\text{diag}(1,1,1,-1)$. Graph states belong to the class of stabilizer states and can thus be equivalently defined by their stabilizers.

Graph states can be manipulated in various ways, among which single-qubit Pauli measurements are of particular interest. Let $P_a^{(\pm)}$ denote the projective measurement of qubit $a$ in basis $P\in\{X,Z,Y\}$ with outcome $+1$ or $-1$. After a measurement, the resulting state on the non-measured qubits, i.e., the state after having traced out the measured qubit, is up to local Clifford operations that depend on the measurement outcome a new graph state. The exact form of the resulting graph state depends on the chosen basis:
\begin{align}
    Z_a^{(\pm)}\ket{G} &=U_{z,a}^{(\pm)}\ket{G-a}\label{eq: Z_measurement_action},\\
    Y_a^{(\pm)}\ket{G} &=U_{y,a}^{(\pm)}\ket{\tau_a(G)-a}\label{eq: Y_measurement_action},\\
    X_a^{(\pm)}\ket{G} &=U_{x,a}^{(\pm)}\ket{\tau_b(\tau_a(\tau_b(G))-a)},\quad b\in N(a)\label{eq: X_measurement_action},
\end{align}
where $\tau_n$ is a local complementation of $G$ with respect to vertex $n$. In simple terms; a $Z$-measurement of a qubit corresponds to vertex deletion of that qubit in the graph; a $Y$-measurement a local complementation with respect to that vertex followed by vertex deletion; a $X$-measurement two local complementations on neighboring vertices, a vertex deletion, and a final local complementation.

No matter the choice of basis, the resulting graph state after a Pauli measurement is given as
\begin{equation}
    \ket{G'}=\left(U_{P,a}^{(\pm)}\right)^\dagger P_a^{(\pm)}\ket{G}.
\end{equation}
The local Clifford operations $\left(U_{P,a}^{(\pm)}\right)^\dagger$ are applied in order to obtain a graph state, and hence they are referred to as correction operations (COs) related to that measurement. The possible COs are listed in Table \ref{tab: Correction_Operations}.

\begin{table}[]
    \centering
    \makegapedcells
    \begin{tabular}{c|c|c|c}
     & $Z$ & $Y$ & $X$\\
     \hline
    $+$ & $I$ & $\sqrt{iZ_{N(a)}}$ & $\sqrt{-iY_b}Z_{N(a)\setminus\{N(b)\cup b\}}$\\
    $-$ & $Z_{N(a)}$ & $\sqrt{-iZ_{N(a)}}$ & $\sqrt{iY_b}Z_{N(b)\setminus\{N(a)\cup a\}}$\\
    \hline
    \end{tabular}
    \caption{Correction operations, $\left(U_{P,a}^{(\pm)}\right)^\dagger$, for Pauli measurements}
    \label{tab: Correction_Operations}
\end{table}

When performing a sequence of Pauli measurement, the COs must be taken into account. This can be done in two ways; apply them, if necessary, before the subsequent measurement or postpone all COs to the end by making appropriate changes to the measurement basis. The latter is generally assumed for MBQN as it removes the need for signaling between each measurement. Thus, letting the $i$-th measurement be performed on qubit $\pi(i)$ with basis $P_i$ results in a graph state on the form
\begin{equation}
    \ket{G'}=\left(U_{P'_n,\pi(n)}^{(\pm)}\right)^\dagger\cdots\left(U_{P'_1,\pi(1)}^{(\pm)}\right)^\dagger(P_n)_{\pi(n)}^{(\pm)}\cdots(P_1)_{\pi(1)}^{(\pm)}\ket{G}.
\end{equation}
Here, $P_i$ denotes the physical measurement basis, i.e., the one used by the measurement apparatus, while $P'$ is the corresponding logical basis that describes the effect of the measurement when postponing all COs to the end. The logical basis and measurement outcome of a measurement differ from the physical measurement basis/outcome if any of the COs related to preceding measurements act non-trivially on the qubit to be measured. Consequently, the logical basis and outcome of the first measurement always coincide with the physical counterparts. Importantly, the required physical measurement basis to obtain a certain logical measurement basis can be computed prior to any measurements being performed due to the commutativity properties of Pauli measurements and the COs (see Table II in \cite{hein2006entanglement}). On the other hand, the logical measurement outcome of a measurement depends on the outcomes of the previous measurements that applies COs on the measured node. To ease readability, the term 'measurement basis' will henceforth refer to the logical measurement basis.

\subsection{Network model}
We consider a decentralized quantum network where nodes can request distribution of entanglement. Following the MBQN framework, a multipartite resource state is initially distributed across the network, which through a sequence of LOCC is transformed to satisfy the task. We emphasize that the aim of this work is to propose a temporal framework that enables causality-preserving scheduling of measurements for a given resource state and network. As a result, we do not explicitly analyze the quantum layer itself, but the implication of the quantum layer on the coordination layer. That is, we do not optimize the resource state design nor the measurement sequence. Furthermore, we do not consider the actual distribution of the resource state, but merely assume it to be pre-shared. Our focus in on resolving the conflicts that arise from the need to access shared resources for carrying out measurements.

For simplicity, we restrict our attention to distributing a single EPR pair between a source $s$ and a receiver $r$. While certain quantum applications involve more general forms of multipartite entanglement, such scenarios can be treated analogously. Restricting attention to single EPR pair distribution suffices to illustrate the causality challenges motivating the temporal framework. Formally, a task is defined as an ordered pair $(s,r)$, where the order merely indicates who initiates the task and thereby the flow of information in the network. In the graph state formalism, an EPR is local Clifford equivalent to a graph consisting only of the edge $(s,r)$. The aim of this paper is therefore to determine a causality-preserving measurement scheduling that partitions the resource state such that $(s,r)$ is a disjoint graph. 

The measurement sequence required to achieve a given task is fixed by the task specification and the network topology, hence we assume this sequence to be known by $s$. What $s$ must still determine is a causality-preserving time-slot assignment for this sequence. Since the scheduling design can be carried out offline before the task is requested, we furthermore assume that $s$ already possesses the resulting schedule. Accordingly, the scheduling computation is not explicitly represented in the temporal framework. A schedule is specified by the measurement basis and the assigned time slot for each measuring node. That is, each node is assigned a pair $(S_i,P_i), S_i\in\mathbb{Z}_+,P_i\in\{I,Z,Y\}$, reflecting the slot to measure in and the logical basis to use; the appropriate physical measurement basis can be calculated by each node from the schedule. Note that we refrain from using $X$-basis measurements. This choice highlights the temporal structure and the notation of causality-preserving measurement scheduling without conflating it with basis-selection optimization.

\subsection{Time-Division Framework}
To preserve the causal ordering required for consistent measurement outcome interpretation, we introduce a time-division model as the fundamental feature of the coordination layer. To expose the source of ambiguity, we assume homogeneous classical propagation delays but heterogeneous quantum hardware with node-dependent measurement times. Consequently, the ambiguity of measurement interpretation arises from quantum heterogeneity, not classical irregularities. This distinction motivates introducing two distinct time units:
\begin{itemize}
    \item Quantum measurement time $T_q$:\\
    The duration for a node to process the received schedule information, configure its hardware and measurement apparatus, perform the assigned single-qubit Pauli measurement, and parse the classical outcome to memory.
    \item Classical communication time $T_c$:\\
    The transmission time for classical information over a single link.
\end{itemize}
The heterogeneity of quantum hardware implies that each node has its own measurement time $T_{q,i}$, which may even be stochastic depending on whether a node is temporarily occupied with other operations. This variability obscures the causal ordering of measurements and, consequently, the interpretation of their outcomes. To eliminate this ambiguity, we introduce a network-aware measurement time $T_q$, defined as the minimal duration in which every node can complete its measurement. Formally, $T_q=\max_i T_{q,i}$.

Based on these two time units, we introduce a dual time-division architecture: quantum slots of length $T_q$, during which a prescribed subset of nodes perform their measurements; and classical slots of length $T_c$, during which nodes can transmit single-hop classical information. The exchanged classical information includes both feedforward regarding the measurement schedule or feedback carrying measurement outcomes. We assume that both classical and quantum slot sequences begin at the instant a task is initialized. Slots of each type occur back-to-back and independently of one another, allowing a node to engage in one classical and one quantum operation concurrently. Operations associated with a given slot are considered to start at the beginning of the slot and to complete by its end.

Introducing time slots may, in general, increase the total execution time of a task by enforcing causality-preserving measurement ordering. However, the imposed temporal structure yields architectural benefits, including unambiguous outcome interpretation, early detection of missing or delayed messages, and supports streaming corrections. In this sense, the temporal overhead introduced by the time-slotting constitutes a deliberate trade-off between execution speed and architectural robustness.

Rather than varying both $T_c$ and $T_q$, we normalize time with respect to the classical slot duration $T_c$. Specifically, we set $T_c=1$ and express all quantum times relative to this unit. This yields three operating regimes: if $T_q\in(0,1)$, classical communication is slower than the measurement time; if $T_q=1$, the two times are equal; and if $T_q>1$, classical communication is faster than measurements. These regimes characterize when measurement timing may compromise causality and motivate the introduction of a temporal coordination layer.

\subsection{Measurement Scheduling Constraints}
To understand exactly when ambiguity arises, we now formalize two types of constraints that define a causality-preserving measurement schedule. First, a node may only measure after knowing which basis to measure in. Second, measurements with unresolved causal dependencies must be separated across slots.

\subsubsection{Feedforward}
In the decentralized network setting considered here where a task is initiated by a source $s$, all other nodes are unknowing of the task -- and thereby the basis to measure in -- until receiving the feedforward information. This feedforward is propagated hop-by-hop over the classical communication network and incurs a delay of one classical slot per hop. Consequently, a node cannot perform its measurement in slot $k$ unless the required feedforward has reached it by the beginning of that slot.

Let $d_s(i)$ denote the number of hops between $s$ and node $i$ in the the classical communication topology. The feedforward constraint can then be formalized as
\begin{equation}\label{eq: Feedforward_constraint}
    S_i\geq \left\lceil\frac{d_s(i)}{T_q}\right\rceil+1,\quad\forall i.
\end{equation}
The feedforward constraint directly links the measurement schedule to the classical communication topology. When $T_q\in(0,1)$, measurement completions occur before a single-hop feedforward can propagate. Consequently, the ordering of measurements is dictated solely by the communications delays. In other words, this regime provides no additional causal ambiguity beyond the $T_q=1$ baseline and therefore does not affect the structure of causality-preserving schedules we aim to study. For this reason, we focus on the regime $T_q\geq1$, where heterogeneous measurement durations can interact non-trivially with classical propagation and lead to the ambiguities that motivate the coordination layer. We emphasize that feedforward is strictly one-way; it propagates outwards from the source along the classical communication paths.

Because the ceiling function in \eqref{eq: Feedforward_constraint} is piecewise-constant, $S_i$ only changes at rational breakpoints of the form
\begin{equation}
    T_q=\frac{d_s(i)}{k},\quad k\in\{1,\ldots,d_s(i)\}.
\end{equation}
Note that $k$ is bounded by $d_s(i)$ due to the restriction to $T_q\geq1$. As a result, it suffices to analyze $T_q$ only at these finitely many breakpoints to characterize the constraint for all possible values of $T_q$. Taking the union of such breakpoints for all nodes that according to the schedule must measure, gives the following set of finitely many breakpoints to consider:
\begin{equation}
    \mathcal{I}=\bigcup_{i}\bigg\{\frac{d_s(i)}{k},\, 1\leq k\leq d_s(i)\bigg\}\cup\{+\infty\}.
\end{equation}
Assuming $\mathcal{I}$ to be ordered increasingly, every possible value of $T_q$ belongs to some interval defined by two adjacent elements in $\mathcal{I}$ on the form $[\mathcal{I}_i,\mathcal{I}_{i+1})$. The feedforward constraint for values in this interval is fully characterized by considering $T_q=\mathcal{I}_i$.

\subsubsection{Adjacency}
Each measurement modifies the underlying resource state topology and generally induces COs on other nodes. The order in which these COs are induced determines the correct interpretation of subsequent measurement outcomes, yet the outcomes themselves carry no information about this order. As a result, if two nodes that can induce COs on one another measure concurrently, the resulting CO sequence becomes ambiguous, leading to measurement outcome interpretation being ambiguous. To prevent such ambiguity, we restrict which nodes may measure in the same slot. In our setting, where we refrain from $X$-basis measurements, Table~\ref{tab: Correction_Operations} shows that COs are only induced on neighbors of the measured node. Thus, two nodes may not both measure in the same slot if they are adjacent in the resource state at the beginning of said slot. This yields the adjacency constraint, formalized as:
\begin{equation}
    S_i=S_j\implies i\notin N(j) \text{ in slot } S_i.
\end{equation}
One might worry that two non-adjacent measuring nodes could both induce COs on a common neighbor $a$, potentially introducing ambiguity. However, when measurements are restricted to the $Z$- and $Y$-bases, the only COs induced on $a$ are drawn from the set $\{I_a,Z_a,\sqrt{iZ_a},\sqrt{-iZ_a}\}$. Since all of these operators commute, the order in which the COs are induced is irrelevant: their combined effect on $a$ is uniquely determined by the measurement bases and outcomes, not by the temporal order of the measurements. Thus, although node~$a$ must eventually learn all the outcomes to determine its local CO, no ambiguity arises from concurrent measurements on its non-adjacent neighbors. Consequently, the adjacency constraint above is sufficient to eliminate all causality-dependent ambiguities.

Note that for each slot, the adjacency constraint corresponds to finding an independent set in the current graph.

\subsection{Network Topology and Resource State}
Under the feedforward and adjacency constraints, designing a causality-preserving measurement schedule is equivalent to iteratively finding independent sets in a graph whose topology evolves each round and whose size gradually decreases. Minimizing the number of slots therefore requires finding a maximum independent set in each round, a problem that is known to be NP-hard even for simple graphs like the 2D cluster \cite{johnson1979computers}. We therefore focus on 1D cluster states, where the optimal solution is known. While schedule optimization is not the focus of this work, using a case with a known optimum allows us to clearly illustrate the temporal framework. To ease the notation, we assume the classical communication topology to align with the initial resource state as in Fig.~\ref{fig: Introduction_Example_Topology}.

\subsection{Streaming Reliability}
Time-division is introduced to ensure interpretability. The adjacency constraints guarantee that once a slot $S$ is complete, the end nodes know exactly which nodes have been measured and can reliably interpret the outcomes from that slot. Only after all (locally relevant) outcomes from slot $S$ are received can they correctly process those from slot $S+1$. In an asynchronous setting, missing feedback is indistinguishable from delayed or lost packets, making it impossible to determine whether the corresponding measurements occurred. Without this temporal structure, streaming becomes unreliable; causal relationships between measurements and their outcomes can be misinterpreted, undermining the correctness of any subsequent processing.

\subsection{Problem Statement}
Having described the model, we can formalize the problem considered to illustrate the temporal model in this work.

We assume the following are given: a 1D cluster resource state; a classical communication topology aligned with the cluster graph; a quantum measurement time $T_q$; and a task $(s,r)$. The goal is to determine a causality-preserving measurement scheduling -- that is, one satisfying the feedforward and adjacency constraints -- that satisfies the task. While the task may impose a quantum application specific latency threshold, and one could consider optimizing the schedule to minimize the number of measurement slots, the focus of this work is not on schedule optimization. Instead, we use the minimum number of rounds as a conceptual reference, which helps illustrate the temporal framework and the effects of the scheduling constraints. Let the minimal number of measurement slots for a given $T_q$ and $(s,r)$ be denoted $T^*(T_q;s,r)$.

\section{Sequential and Parallel Measurements}\label{sec: Extreme_Cases}
In this section, we illustrate the notion of time-division and measurement scheduling by examining two extreme cases. We consider a network of 6 nodes and determine the measurement schedule for the task $(s=1,r=5)$. To satisfy this task, all other nodes in the network must perform a measurement. In the first case, $T_q=1$, corresponding roughly to the general MBQN framework where each measurement completes before the next begins. In the second case, $T_q\geq5$, allowing all nodes to, in principle, measure concurrently. These two cases correspond to opposite extremes: in the first, measurements are fully feedforward-constrained, resulting in a sequential schedule; in the second, measurements are fully adjacency-constrained, allowing for fully parallel execution to the extent permitted by the constraints.

\subsection{Sequential}
In the fully sequential case, $T_q=1$, the classical communication slots align perfectly with the quantum measurement slots. Consequently, when a node receives the feedforward information, it must immediately perform two actions; relay the feedforward information along the network path and measure its qubit in the basis determined by the schedule. In this scenario, the durations of these operations are equal, whence a measurement always completes before the next node along the path can begin its measurement.

The usual ``measure-as-soon-as-possible'' assumption in MBQN therefore applies without risk of  ambiguity in measurement interpretation. Each node measures in the earliest slot allowed by the feedforward constraints, which for this network and task are $S\geq[1,2,3,4,5,6]$. Since the last measurement (node~6) is performed in the sixth slot, the sequential schedule for this network and task yields $T^*(1;1,5)=6$. One schedule achieving this is
\begin{equation}
    (S,R)=[(1,I),(2,Y),(3,Y),(4,Y),(5,I),(6,Z)].
\end{equation}
An analogous analysis applies for $T_q\in(0,1)$. While the earliest slots allowed by the feedforward constraints are slightly adjusted as at least one quantum slot change occurs within each classical slot, the overall structure of the analysis remains unchanged. For example, $T^*(1/2;1,5)=11$, as only every other quantum slot can be used for measurements due to the slow feedforward propagation.

\subsection{Parallel}\label{sec: Extreme_Cases_Parallel}
In the fully parallel case, $T_q\geq5$, determining an optimal schedule is less trivial because the adjacency constraints introduce interdependencies between measurements. The rapid propagation of feedforward information makes outcomes of concurrent measurements potentially ambiguous if adjacency constraints are violated. Here, the feedforward constraints give $S\geq[1,2,2,2,2,2]$, so reasoning about adjacency is crucial. For example, at most two of the three nodes between node $s$ and $r$ can measure simultaneously; consequently, at least one node must delay its measurement by at least one round to satisfy the adjacency constraints. One optimal scheduling for this case is
\begin{equation}
    (S,P)=[(1,I),(2,Y),(3,Y),(2,Y),(2,I),(2,Z)].
\end{equation}
From this schedule, we see that $T^*(5; 1,5)=3$. This is optimal because not all nodes between $s$ and $r$ can measure concurrently in the second slot. By exploiting the compatibility of measurements on non-adjacent nodes, the number of measurement slots can be significantly reduced, here by $50\%$, compared to fully sequential case.

We also note that slot assignments are generally not interchangeable. For example, one might think that as the assignment $S_2=2,S_3=3,S_4=2$ is valid, then the ``opposite'' assignment $S_2=3,S_3=2,S_4=3$ would also be valid, since the number of nodes per slot is the same, and only non-adjacent nodes are assigned the same slot. However, measuring $Y_3$ in slot two links nodes~2 and 4, making them ineligible to measure in the same slot due to the adjacency constraints.

\section{Intermediate Quantum Slot Regimes}\label{sec: Scheduling_Heuristics}
While the two extreme cases illustrate the concept of measurement scheduling, each considers only one type of constraint while the other is trivially satisfied. For most values of $T_q$, both feedforward and adjacency constraints arise non-trivially. In this section, we extend the intuition obtained from the extreme cases to arbitrary values $T_q$ and analyse the optimal scheduling under the considered model.

Consider an arbitrary task $(s,r)$. To simplify notation, we assume $s<r$ without loss of generality due to the symmetry of the network. Satisfying this task, requires two steps: first, isolating a path between nodes $s$ and $r$ from the rest of resource state; second, converting this path into a single long edge $(s,r)$. The first step is achieved by performing $Z$-measurements on the outer neighbors $s-1$ and $r+1$, if they exist. The second step is satisfied by $Y$-measurements on the inner nodes $s+1,\ldots,r-1$, if they exist. Importantly, these two steps can be performed independently. This independence is also evident in the extreme cases: in the sequential case, the path is isolated after it is converted to the edge $(s,r)$; in the parallel case, the steps occur simultaneously. Since each task requires only one type of measurement on the outer nodes and one type on the inner nodes, scheduling reduces to assigning a measurement slot to each node.

\subsection{Outer Neighbor Scheduling}
We first consider measurements on the outer neighbors. Because these can be performed independently of measurements on both the inner nodes and each other, we always schedule outer nodes to measure as early as possible. The feedforward constraint for the outer node adjacent to the source is $S_{s-1}=\lceil 1/T_q\rceil+1=2$ for any value of $T_q$. For the outer node adjacent to the receiver, the constraint is $S_{r+1}=\lceil (r-s+1)/T_q\rceil+1$, which can take values in $S_{r+1}\in\{2,\ldots,r-s+2\}$. This implies that, if both outer nodes exists, the receiver-adjacent node is always the bottleneck, as expected. The source-adjacent node always measures in slot $S_{s-1}=2$ -- the earliest possible slot -- and therefore it can never be the unique bottleneck unless it is the only measuring node in the network.

We will now consider the measurements on the inner nodes. First, we extend the sequential scheduling to any $T_q$-value as this can be considered a worst-case scheduling. Naturally, one can delay all measurements in a scheduling by one slot and obtain a new scheduling with worse performance; but we refrain from considering such naive solutions.

\subsection{Sequential Scheduling}
Enforcing a sequential measurement schedule, regardless of the feedforward propagation speed, is the simplest possible design. The schedule can be constructed by assigning each inner node $s+i$ for $i=1,\ldots,r-s-1$ to measure in slots $S_{s+i}=i+1$. With this assignment, both feedforward and adjacency constraints are trivially satisfied: the feedforward constraint of node $s+i$ is satisfied by the restriction $T_q\geq1$; the adjacency constraints are ensured by assigning only a single measurement per slot. As the final inner node $r-1$ is assigned slot $S_{r-1}=r-s$, the sequential schedule requires $r-s$ quantum slots to complete the inner node measurements.

If the receiver has an outer neighbor, its assigned slot falls in $S_{r+1}\in\{2,\ldots,r-s+2\}$, which may at worst add two extra measurement slots to the schedule. Only when $T_q<(r-s+1)/(r-s-1)$ the total number of slots will increase.

\subsection{Parallelized Scheduling}
Sequential schedules are generally not optimal, as some combination of tasks and values of $T_q$ allow for partial parallelism. In particular, if the feedforward propagation is sufficiently fast, several nodes can be assigned to the same slot. When relying solely on $Y$-measurements, determining the maximum number of nodes that can be assigned to a given slot is equivalent to finding the maximum independent set on the sub-path of nodes that have received the feedforward information. For a 1D cluster graph, this is straightforward; each slot can be assigned approximately half of the feedforward-eligible, yet unassigned nodes. To guarantee optimality, it is important to assign as many nodes as possible per slot, as highlighted in Sec.~\ref{sec: Extreme_Cases_Parallel}. To formalize this, we define the following sets for the inner nodes. Let $E_k(T_q)$ denote the set of nodes that are feedforward-eligible at the start of slot $k$, i.e., nodes that have received the feedforward information:
\begin{equation}
    E_k(T_q)=\{i\in(s,r)\,|\,d_s(i)\leq kT_q\}.
\end{equation}
Note that $T_q\geq1$ implies $E_1(T_q)\subsetneq E_2(T_q)\subsetneq \cdots\subsetneq E_n(T_q)=E_{n+1}(T_q)=\cdots=\{s+1,\ldots,r-1\}$ where $n=(r-s-1)/T_q$. Hence, at least one new node becomes eligible in each successive slot until all nodes are eligible by slot $(r-s-1)/T_q$.

The adjacency constraints restrict which of the eligible nodes can be measured in the same slot. We  define $U_k$ as the set of feedforward-eligible yet unassigned nodes at the beginning of slot $k$:
\begin{equation}
    U_k=\{i\in E_k(T_q)\,|\,S_i\nleq k\}.
\end{equation}
Designing the optimal schedule then becomes a problem of finding a maximum independent set of $U_k$ in each round. For a 1D cluster, a constructive solution is to assign $\lceil\abs{U_k}/2\rceil$ nodes per slot. The procedure for this assignment is given in  Alg.~\ref{alg: Parallelized_Scheduling}.

\begin{algorithm}
    \caption{Parallelized Scheduling Design}
    \label{alg: Parallelized_Scheduling}
    \begin{algorithmic}[1]
        \REQUIRE Task $(s,r)$, measurement time $T_q$

        \STATE In round $k\geq1$, consider the feedforward-eligible yet unassigned nodes $U_k$.
        \STATE Pick the first element in $U_k$. If $U_k$ is empty, all nodes have been assigned and the procedure stops.
        \STATE Assign slot $S=k+1$ to the chosen node and every odd-indexed element in $U_k$.

        \ENSURE Slot assignments $S$ for all nodes between $s$ and $r$ for the given $T_q$.
    \end{algorithmic}
\end{algorithm}
This procedure satisfies the feedforward constraints because only feedforward-eligible nodes are assigned in each round, and the adjacency constraints are satisfied by assigning only every odd-indexed element of $U_k$. Each round assigns $\lceil\abs{U_k}/2\rceil$ nodes, leaving the remaining $\lfloor\abs{U_k}/2\rfloor$ nodes as unassigned for the next round. Using this recursive procedure, the total number of measurement slots can be determined, although no analytical expression generally exists. In the extreme cases, this reduces to the previous results: for $T_q=1$, only a single new node is feedforward-eligible in each slot ($\abs{U_k}=1$), recovering the sequential case with $T^*(1;s,r)=r-s$ rounds. For $T_q\gg1$, full parallelism is possible, yielding total slot number $T^*(\gg1;s,r)=\lfloor\log_2(r-s-1)\rfloor+2$ by recursively removing a maximum independent set from the graph. Therefore, for arbitrary values of $T_q$, the total number of slots satisfies $T^*(T_q;s,r)\in[\lfloor\log_2(r-s-1)\rfloor+2,r-s]$.

\section{Simulations}\label{sec: Results}
To illustrate the behavior of the measurement scheduling framework, we numerically evaluate the total number of measurement slots $T^*$ for different tasks and quantum measurement times $T_q$. Due to network symmetry, we consider tasks defined by a source-receiver pair separated by a distance $D=r-s$ along a 1D cluster. By systematically varying $D$ and $T_q$, we explore how the scheduling transitions between fully sequential and fully parallel regimes.

\begin{figure}
    \centering
    \includegraphics[width=.95\linewidth]{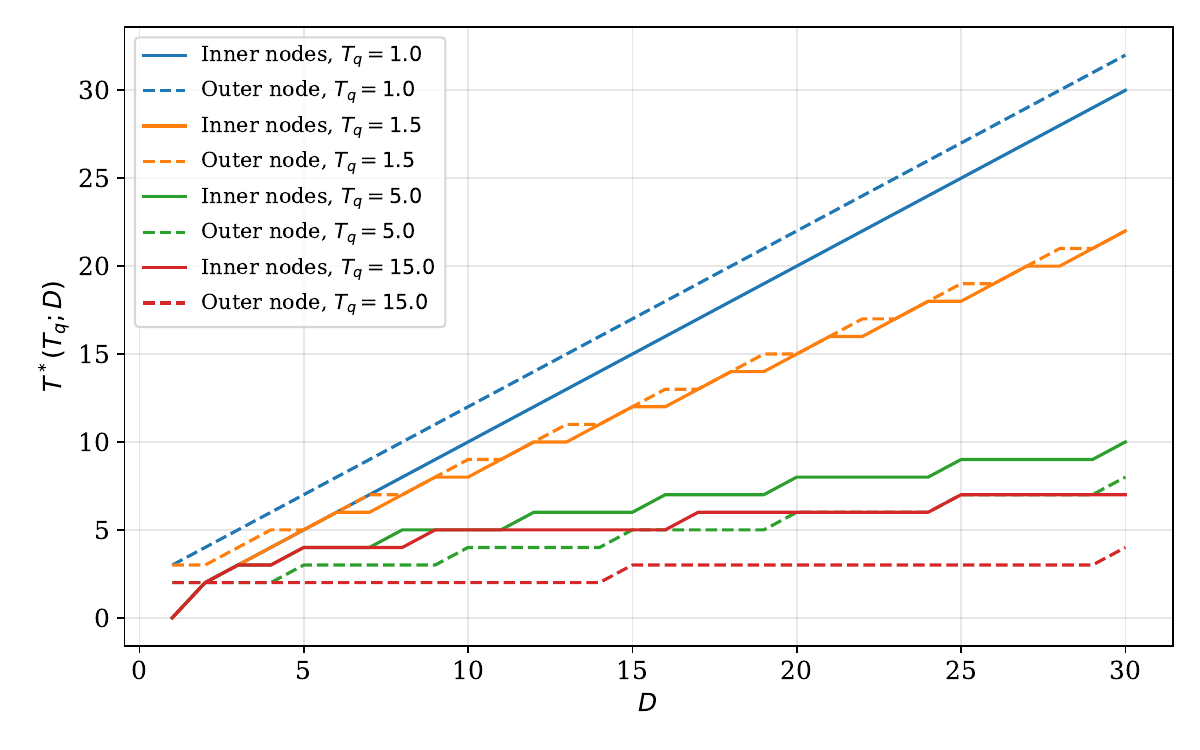}
    \caption{Total number of measurement slots $T^*$ as a function of distance $D$ between source and receiver for selected values of $T_q$.}
    \label{fig: Slots_varying_D}
\end{figure}

The distance $D$ does generally not suffice to determine the number of slots to satisfy a task as it depends on whether the source and/or receiver has outer neighbors. In Fig.~\ref{fig: Slots_varying_D}, we plot $T^*$ as a function of $D$ for several representative values of $T_q$. To highlight the difference between inner and outer nodes, we plot their $T^*$-values separately for each value of $T_q$. The topmost (blue) corresponds to the sequential regime where measurements are fully feedforward-constrained, while the bottom curve (red) corresponds to the parallel regime where measurements are fully adjacency-constrained (at least for $D\leq
T_q$). Intermediate curves illustrate the gradual transition from sequential to parallel schedules as $T_q$ increases. The plot illustrates that for a fixed $T_q$, $T^*$ generally increases as $D$ grows. Interestingly, the schedule for inner nodes is superior to that of outer nodes for low $T_q$, but becomes inferior for large $T_q$. This coincides with feedforward being the bottleneck for low values of $T_q$, while the bottleneck for high values is the adjacency constraints. In particular, the parallel schedules have better performance than sequential in any case for large distances, but also for short distances when $T_q$ is sufficiently high.

\begin{figure}
    \centering
    \includegraphics[width=.95\linewidth]{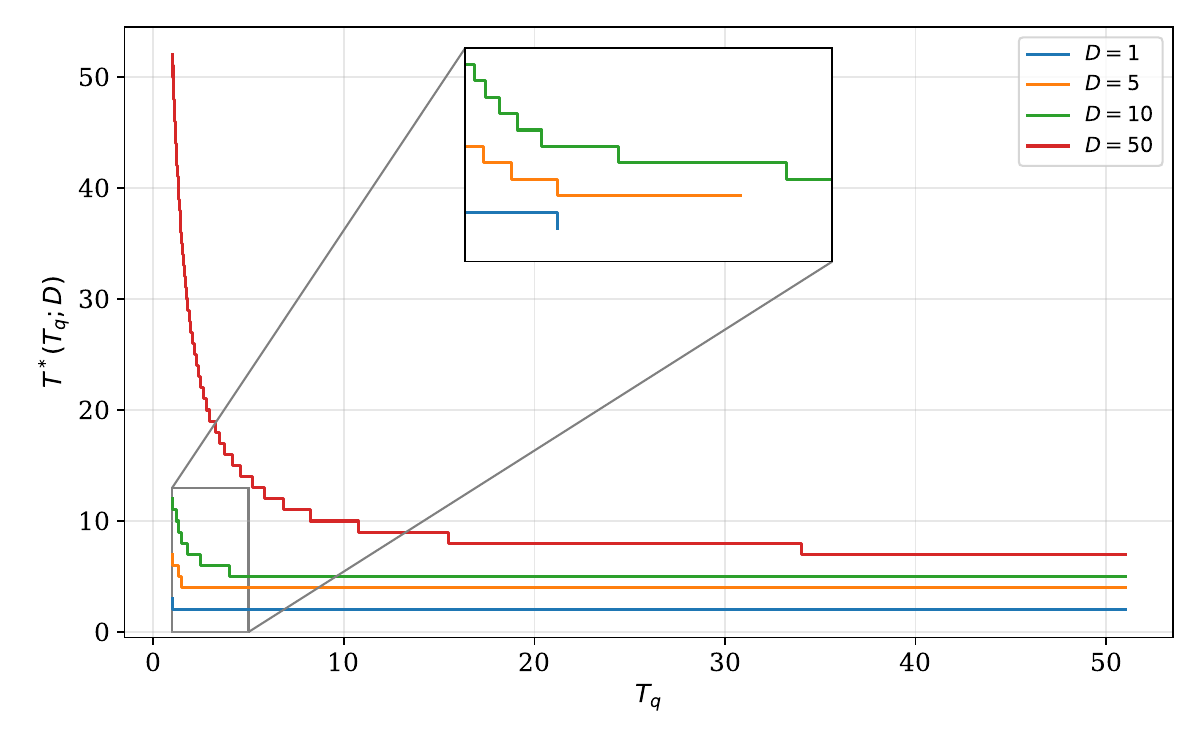}
    \caption{Total number of measurement slots $T^*$ as a function of $T_q$ for several fixed distances $D$. The inset highlights the behavior for small $T_q$ and small $D$, showing that near-optimal schedules are achieved even at modest feedforward speeds.}
    \label{fig: Slots_varying_Tq}
\end{figure}

In Fig.~\ref{fig: Slots_varying_Tq}, we plot $T^*$ as a function of $T_q$ for several fixed distances $D$. Here, we take $T^*$ to be the max of the inner- and outer-node schedules, corresponding to the case where the task is not located at the edge of the network. As expected, $T^*$ decreases rapidly as $T_q$ increases; larger distances exhibit both a higher value at $T_q$ and a higher asymptotic limit as $T_q$ tends to infinity. Conceptually, this reflects the fact that parallel schedules can achieve exponentially fewer slots than purely sequential ones. The inset highlights an important observation; even modest values of $T_q$ are sufficient to achieve the optimal number of slots for short medium-range tasks. Beyond this regime, further increases in $T_q$ provide no additional improvement. This demonstrates that near-optimal scheduling does not require unrealistically fast quantum measurements, but instead emerges naturally once feedforward propagation becomes moderately fast.

\section{Conclusion}\label{sec: Conclusion}
We introduced a time-division measurement framework for measurement-based quantum networks, motivated by the need to eliminate ambiguity in interpreting measurement outcomes across heterogeneous quantum hardware and hop-by-hop classical communication. By enforcing slotted measurement schedules subject to simple feedforward and adjacency constraints, the framework guarantees a consistent causal ordering of outcomes and enables reliable streaming of Pauli-frames, thereby reducing both classical memory requirements and post-processing latency. 

We formalized the model and analysed how measurement durations, classical propagation speeds, and network topology jointly shape feasible schedules. Practical schedule designs were presented to illustrate slot assignment and to clarify the transition between sequential and parallel regimes.

While optimal scheduling can be computed for small or uniform networks, the framework generalized naturally to large, heterogeneous settings. Future work includes tighter optimality bounds, extending the analysis to richer topologies, and incorporating fault-tolerant requirements to further connect classical networking concepts with distributed quantum information processing.

\bibliographystyle{IEEEtran}
\bibliography{bibliography}
\end{document}